\documentclass[runningheads]{llncs}

\usepackage[T1]{fontenc}
\usepackage{hyphenat}

\usepackage{graphicx}

\begin{document}

\title{Pluri\hyp{}perspectivism in Human-robot Co-creativity with Older Adults }
%
%
\author{Marianne Bossema\inst{1,2} \orcidID{0000-1111-2222-3333} \and
Rob Saunders\inst{2} \orcidID{0000-1111-2222-3333} \and
Aske Plaat\inst{2} \orcidID{0000-1111-2222-3333} \and
Somaya Ben Allouch\inst{1,3} \orcidID{0000-1111-2222-3333}}
\authorrunning{M. Bossema et al.}
%
\institute{Amsterdam University of Applied Sciences, The Netherlands \and Leiden University, The Netherlands
\and University of Amsterdam, The Netherlands}
\maketitle              
\begin{abstract}
This position paper explores pluri\hyp{}perspectivism as a core element of human creative experience and its relevance to human–robot co-creativity. We propose a layered, five-dimensional model to guide the design of co-creative behaviors and the analysis of interaction dynamics. This model is based on literature and results from an interview study we conducted with 10 visual artists and 8 arts educators, examining how pluri\hyp{}perspectivism supports creative practice. The findings of this study provide insight how robots could enhance human creativity through adaptive, context-sensitive behavior, demonstrating the potential of pluri\hyp{}perspectivism. This paper outlines future directions for integrating pluri-perspectivism with vision-language models (VLMs), to support context sensitivity in co-creative robots.

\end{abstract}
\section{Introduction}
This study is part of a broader project investigating how robots and artificial intelligence (AI) can support and enhance creative experiences for older adults. Participatory arts have been shown to promote cognitive health, lifelong learning, and social connectedness in later life~\cite{fancourt2017arts,fancourt2019evidence,groot2021value,lewis2024arts}. Technology for creativity support can enhance self-expression~\cite{macritchie2023use}, while robots can uniquely contribute through embodied, co-present, social  interaction~\cite{tanner2023older,breazeal2019designing}. Yet, the potential of robotic creativity support for older users remains underexplored~\cite{bossema2023human}. Integrating VLMs into robots enables contextual understanding, goal planning, and natural interaction~\cite{zhang2023large}. In creative settings, this allows for flexibility and accessible conversational interfaces, particularly valuable for target groups with diverse goals and digital skills, such as older adults. In a participatory study offering a course on ''Drawing with Robots'' for this target group, we found that participants appreciated a VLM-enhanced robot’s vision and language abilities, but noted a lack of sensitivity to the creative context, including their artistic intentions~\cite{bossema2025llm}.  

When acting in the real world, robotic agents (intentional, adaptive, autonomous~\cite{wooldridge1995intelligent}) require an understanding of the context~\cite{vossen2019modelling,plaat2025agentic}. VLMs can support robotic agents by aligning visual inputs with natural language, enabling them to interpret scenes and engage in conversation~\cite{nwankwo2024conversation}. VLMs still require domain-specific guidance, however, to be effective in embodied, situated tasks~\cite{tsimpoukelli2021multimodal,bubeck2023sparks}. Human feedback can provide guidance~\cite{thomaz2008teachable}, and recent studies show how VLM-enhanced conversation can facilitate the exchange of contextual information and improve robot task execution~\cite{nwankwo2024conversation}. Yet, important questions remain about what contextual information is needed for creative collaboration, and how this can best be exchanged within a creative process. We propose a pluri-perspectivist model as a contextual map defining a multidimensional co-creative space. Pluri-perspectivism, or exploring and integrating multiple perspectives, is pivotal for collaboration~\cite{clark1996using,mahyar2014supporting} and for creativity~\cite{gluaveanu2014creativity}. The pluri-perspectivist model can be used as an internal schema to guide a VLM-enhanced robot~\cite{lee2021dialogue,zhang2023sgp}, improving robot context sensitivity. Technical approaches are further investigated in Section~\ref{section:related-works}. In human-robot co-creativity, a VLM-enhanced robotic agent could explore and integrate perspectives through both conversation and action, for example by asking about user preferences, detecting turn-taking rhythms, offering suggestions, or synchronizing its actions. In such a setting, the pluri-perspectivist model serves as a tool for structuring contextual exploration in different dimensions. It offers a way to guide attention, interaction, and interpretation in VLM-enhanced robots by informing structured prompts and examples. Using the model as a contextual map may help identify how a robot could act as a co-creative agent, and what forms of perspective taking and offering are meaningful and desirable in a co-creative process.

The goal of this study is to introduce a pluri-perspectivist model that can be used as a contextual map, guiding human-robot co-creativity. We review related work on perspective sharing in creative human-robot interaction (HRI) contexts, and explore technical approaches to assess potential implementation of the model in a schema-guided~\cite{lee2021dialogue,zhang2023sgp}, VLM-enhanced robot. In addition, we investigate artistic exploration in practice through interviews with experts (artists and art teachers) to validate the model and to collect meaningful examples. The model offers a theoretically grounded, practical framework for co-creative HRI. Ongoing work involves deploying the model with a VLM-enhanced robot to assess its applicability in HRI. We view this preliminary study as a necessary foundation for future implementation and empirical testing in HRI, whereby mapping a domain-specific context contributes to the broader goal of enhancing robotic agents’ contextual understanding.

Section~\ref{section:background} presents background literature. Section~\ref{section:proposition} outlines the proposition and model. Section~\ref{section:related-works} draws from related work on robot behaviors and technical approaches, and Section~\ref{section:interviews} describes the interview study. Sections~\ref{section:discussion} and~\ref{section:conclusion} discuss findings and conclude with directions for future research.

\begin{figure}[t] 
\centering
\includegraphics[width=\linewidth]{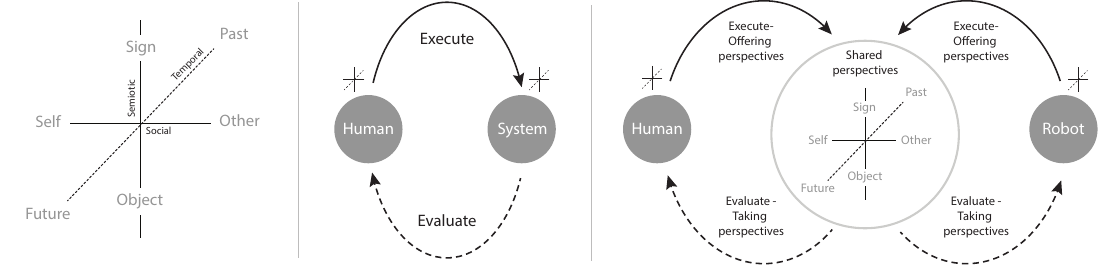} 
\caption{Left: The model of creatogenetic differences by Glaveanu \& Gillespie~\cite{gluaveanu2014creativity},  Center: The HCI Interaction Cycle by Norman~\cite{norman1986cognitive}; Right: Integration of model and interaction cycles in human-robot co-creativity}
\label{fig:gulfs}
\end{figure}

\section{Background}\label{section:background}

\subsubsection{Creative Experiences and Pluri\hyp{}perspectivism}--- The main question of our broader research project is how robots and AI can support and enhance creative experiences for older adults. We adopt Glăveanu \& Beghetto’s~\cite{gluaveanu2021creative} definition of creative experiences as \textit{``novel person-world encounters grounded in meaningful actions and interactions, marked by openness, non-linearity, pluri-perspectives, and future orientation.''} The researchers suggest that a key principle of creative experiences is pluri\hyp{}perspectivism, the active search for and engagement with different viewpoints. Glăveanu argues that creativity not only results in difference, but also originates in it~\cite{gluaveanu2023difference}. This means that exploring and integrating difference in a multidimensional possibility space is central to creativity. In ``Creativity out of Difference'', Glăveanu \& Gillespie present three creatogenetic differences---between self and other, between sign and object, and between past and future---modeled into social, semiotic and temporal perspectives (Figure~\ref{fig:gulfs}, left) \cite{gluaveanu2014creativity}. Expanding on sociocultural psychological models, the authors argue that all forms of creativity arise from these dynamic self-other-world-sign relationships, shaped through interactions over time. These social, semiotic and temporal perspectives are the foundation of our pluri-perspectivist model for human-robot co-creativity.

\subsubsection{Pluri\hyp{}perspectivism in Creative Collaboration}---Boden~\cite{boden2004creative} defines creativity as generating ideas or artifacts that are both novel and appropriate. In creative collaboration, partners introduce and evaluate novelty and appropriateness together. While the dynamics depend on roles and distribution of agency, this requires exchanging perspectives with regards to 1) the creative task, and 2) the collaborative process. Exchanging perspectives involves perspective taking and perspective offering. Regarding the creative goal, perspective taking is about perceiving and evaluating the creative possibility space, while perspective offering means reshaping that space~\cite{gluaveanu2014creativity}. Regarding the collaborative goal, perspective taking is about recognizing others' viewpoints~\cite{galinsky2005perspective}, while perspective offering may involve suggesting a way of working. Exchange of perspectives happens in a close loop using multiple modalities. In Human–Computer Interaction (HCI), Norman’s interaction cycle~\cite{norman1986cognitive} reflects a similar close loop of perception, evaluation and execution. Users engage in perspective taking towards a system by evaluating system feedback based on their expectations and offer new perspectives by executing actions. Figure~\ref{fig:gulfs} (center) illustrates this unidirectional loop between human and system. In collaborative tasks, however, this process is bidirectional: agents jointly act and evaluate within a process that can be understood by sharing perspectives in different dimensions. This is shown on the right side of the figure, where human and robot share social, semiotic, and temporal perspectives, as defined by Glaveanu and Gillespie's model of three creatogenetic differences~\cite{gluaveanu2014creativity}. In creative collaboration, a pluri-perspectivist model can help to structure the dynamic interplay of perspective taking and perspective offering in multiple dimensions (Figure~\ref{fig:gulfs}, right). In human-robot co-creativity, it can guide a VLM-enhanced robot in exchanging perspectives with regard to the creative task, and to the collaborative process. 

\begin{figure}[t] 
\centering
\includegraphics[width=0.85\linewidth]{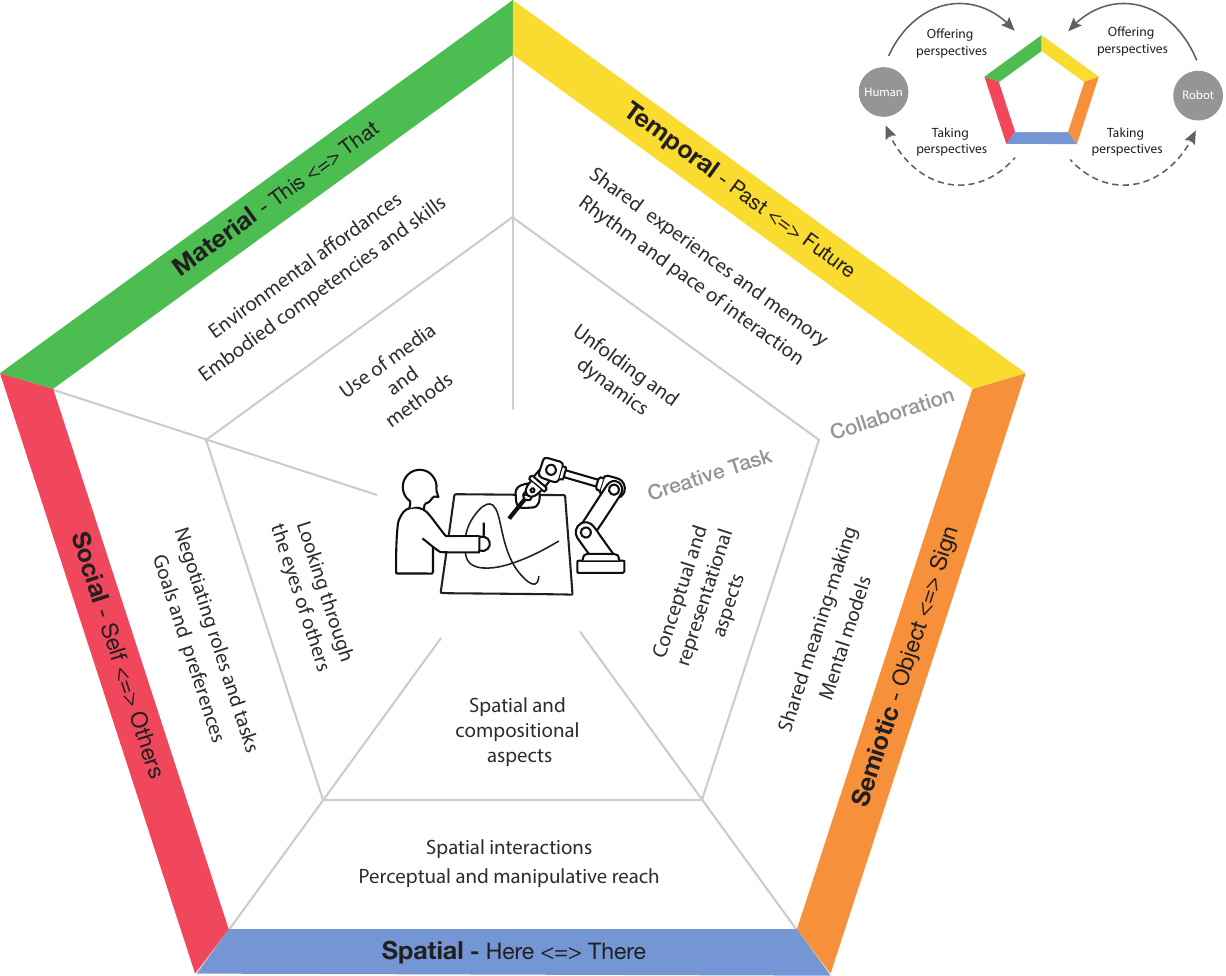} 
\caption{The two-layered five-dimensional model of pluri\hyp{}perspectivism (left) for interaction cycles in human-robot co-creativity (top-right).}
\label{fig:pentagram}
\end{figure}

\section{Proposition}\label{section:proposition}
Taking Glăveanu \& Gillespie's three-dimensional creatogenetic model~\cite{gluaveanu2014creativity} as a starting point, we propose a five-dimensional contextual map for human-robot collaborative creativity that can guide a VLM-enhanced robots in co-creative tasks. 

\subsubsection{Five Dimensions}---We build on the three creatogenetic differences (Figure~\ref{fig:gulfs}, left) and introduce `material' and `spatial' as additional, context-specific dimensions. We include these dimensions based on the theory of participatory sense-making by De Jaegher and Di Paolo~\cite{de2007participatory}, that describes how meaning is co-created through embodied, dynamic interactions between agents. In co-creative contexts, meaning-making emerges not only through linguistic or symbolic exchange but also through embodied interaction and engagement with the material environment. Including material and spatial dimensions in the model allows us to capture the dynamic, situated nature of creativity that unfolds through embodied action, perceptual alignment, and interaction with physical media. Several related works underpin the importance of spatial and material perspectives. Saunders et al.~\cite{saunders2013evaluating} demonstrate how sharing a physical environment enables richer social and cultural exchange. Guckelsberger et al.~\cite{guckelsberger2021embodiment} demonstrate that embodiment influences human perception of robot creativity and intentionality, and Weinberg et al.~\cite{weinberg2020social} highlight how robot embodiment enhances collaboration, turn-taking, and engagement in social and musical contexts. By mapping the five dimensions -social, material, temporal, semiotic and spatial we can investigate in what ways robots as material agents can contribute to human-robot co-creativity. Definitions of the five dimensions are listed in Table~\ref{tab:coding_scheme}.

\subsubsection{Two Layers}--- Co-creative processes involve actions related to 1) coordination and 2) task execution, and the model adopts a two-layered structure to support both. To conceptualize the distinction between the two layers, we draw on Clark's theory of joint action~\cite{clark1996using}, which differentiates between task-level actions---contributions aimed at accomplishing the task---and coordination-level actions, that facilitate the collaborative process itself. This theoretical framing aligns with our model: The inner layer supports the creative task, while the outer layer encompasses collaborative actions that help organize, negotiate, and guide the co-creative process. Figure~\ref{fig:pentagram} shows the five perspectives in a two-layered model.

\subsubsection{Speculative Scenario}
\begin{figure}[t] 
\centering
\includegraphics[width=\linewidth]{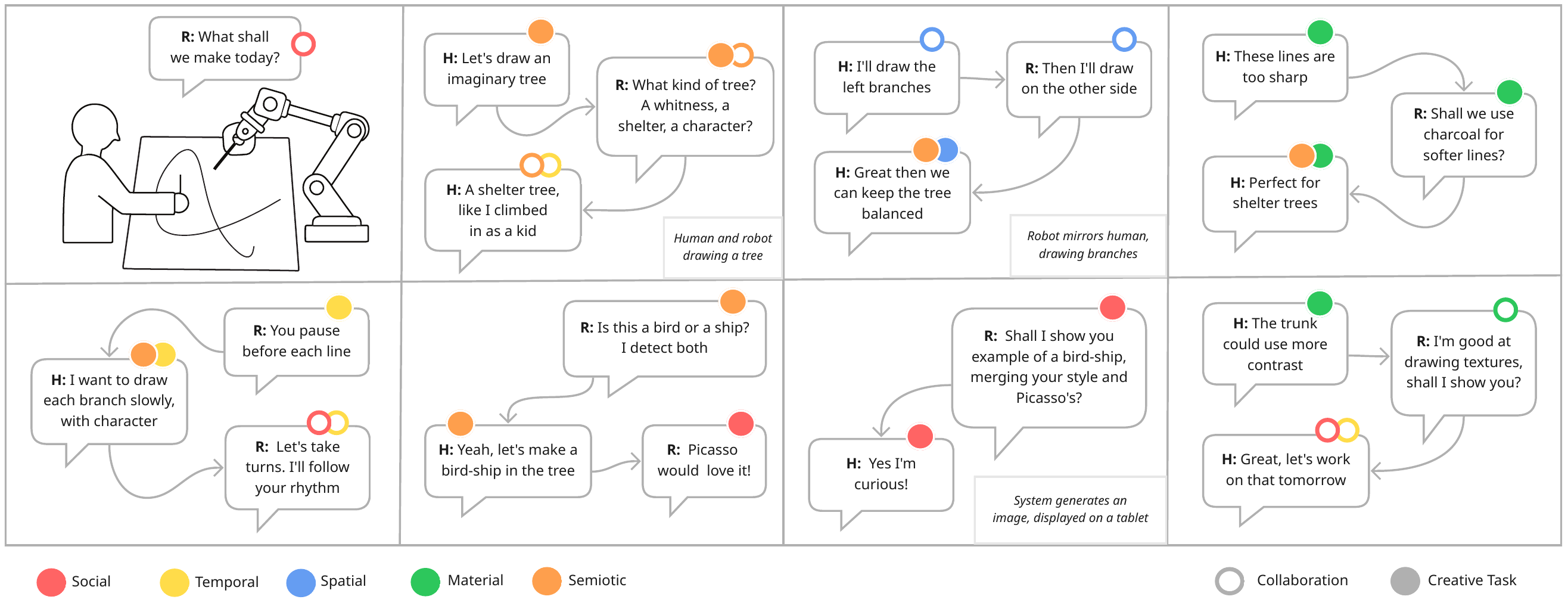} 
\caption{pluri\hyp{}perspectivism in a speculative scenario of human-robot drawing}
\label{fig:speculative-scenario}
\end{figure}

--- We first tested the proposition in a speculative scenario of human-robot drawing, visualized in figure~\ref{fig:speculative-scenario}. The scenario illustrates how pluri\hyp{}perspectivism could manifest itself in human-robot co-creativity. For example, the dialogue between human (H) and robot (R) unfolds through spatial actions (\textit{``I’ll draw mine to the right''}), material choices (\textit{``Shall we use charcoal?''}), and embodied turn-taking (\textit{``Let’s take turns, I’ll follow your rhythm''}). The scenario reveals that a single expression may engage multiple dimensions simultaneously, enabling fluid perspectival shifts that guide creative exploration. The scenario also reveals structural dependencies between perspectives, suggesting a staged or layered unfolding of interaction. Role negotiation, for instance, often precedes task execution, aligning with Clark’s theory of joint action~\cite{clark1996using} and Engeström’s activity theory~\cite{engestrom2014learning}, where coordination-level actions enable goal-directed behavior. Collaboration-level perspectives may often precede and structure the unfolding of creative task-level perspectives.

\section{Related Works}\label{section:related-works}
To validate the model and inform future implementation with a VLM-enhanced robot, we investigate perspective sharing robot behaviors that support creativity, and possibilities for implemention in VLM-enhanced robots.

\subsubsection{Perspective Sharing Behaviors in Creative HRI}---Research shows that robots can support creativity through a range of perspective sharing behaviors across the dimensions of the pluri\hyp{}perspectivist model. In the \textbf{social} dimension, robots offered perspectives by adopting expressive behaviors that enhanced creativity~\cite{ali2021social,gomez2021robot,rond2019improv}. Fucinato et al.~\cite{fucinato2023charismatic} showed that charismatic robot speech enhanced team creativity. Robots have also acted as creative instructors, using guided demonstrations to stimulate children's artistic output~\cite{ali2019can,elgarf2022creativebot}. Social alignment in robots, through perspective-taking~\cite{almeida2023would} or personality matching~\cite{whittaker2021designing}, has been shown to enhance user preference and prosocial engagement. In the spatial dimension, Lee~\cite{lee2024adversarial} showed that a robot can elicit creativity by disrupting physical arrangements, where changes in tool placement led to new compositions. Robots have demonstrated spatial perspective-taking by interpreting deictic references through language and body posture relative to the user’s viewpoint~\cite{pandey2011towards}, and by using gaze and gesture to promote shared attention~\cite{zhao2015people}. The \textbf{semiotic} dimension includes robots suggesting reinterpretations of creative work. Hu et al.~\cite{hu2021exploring} and Lin et al.~\cite{lin2020your} showed that robots manipulating visual elements or suggesting alternate meanings can prompt conceptual shifts. Several studies illustrate overlaps e.g., \textbf{social and semiotic} perspectives in Tseng's~\cite{tseng2024two} study, where personas generated by a large language model provided conversations inspiring creative meaning-making (semiotic). In Rond's study of improvisational robots~\cite{rond2019improv}, \textbf{spatial and semiotic} perspectives overlap, showing that robot movement can serve as both spatial and symbolic prompts in improvisation. Together, these studies demonstrate how robots can engage in multidimensional perspective sharing to support creativity and collaboration.

\subsubsection{Implementing pluri-perspectivism with a VLM-enhanced robot}---Recent advances in VLMs offer a technical foundation for implementing perspective sharing behaviors in co-creative robots, allowing for guidance through schema-guided prompting~\cite{lee2021dialogue,zhang2023sgp}. Thereby, an internal schema helps to organize what information to pay attention to, how to interpret it, and how to respond. Studies in schema-guided dialogue~\cite{lee2021dialogue,zhang2023sgp} and multimodal instruction-following~\cite{tsimpoukelli2021multimodal} demonstrate how prompt structure and multimodal grounding enable dynamic, context-sensitive interaction. These approaches build on in-context learning~\cite{dong2022survey}, where a pre-trained model makes predictions based on a few structured examples, a form of few-shot learning that does not require retraining~\cite{brown2020language,plaat2025agentic}. Our pluri-perspectivist model can be used for schema-guided prompting, while additional examples, derived from artistic practice, can be used for few-shot learning. In this setting, the model serves as a map for exploring and shaping a co-creative possibility space. For instance, few-shot prompts can condition a VLM to engage in role-playing (social), reinterpret artwork (semiotic), or suggest the exploration of other tools (material). Behaviors may be adapted to user preferences derived from conversation, enabling more personalized interaction. Some dimensions align well with schema-guided prompting, while others require more than text-based and visual prompts. VLMs can support social and semiotic dimensions through multimodal understanding, however they struggle to induce temporal or spatial structure ~\cite{yang2025llm,imam2025can}. Temporal reasoning often depends on structured representations like event schemas~\cite{xiong2024large}, and spatial reasoning remains limited without scene graphs~\cite{yang2025llm} or fine-tuning on spatial data~\cite{chen2024spatialvlm}. In embodied co-creative tasks, such as human–robot drawing, we can extend a VLM with additional modules e.g., for human action recognition~\cite{emanuel2021snapshot}, tracking the drawing process using OpenCV~\cite{bradski2000opencv}, and collecting temporal data on rhythms and pacing using custom scripts. The VLM can serve as a central reasoning engine, whereby the pluri-perspectivist model can guide integration, facilitating multidimensional exploration.

\begin{table}
    \caption{Perspectival Dimensions as Main Coding Categories}
    \begin{tabular}{ |p{1.7cm}|p{10cm}| }
    \hline
    Social & Difference between self and others - Considering or responding to different goals, preferences, expressions. \\
    \hline
    Spatial & Difference between here and there -  Engaging with different physical or visual positions, orientations, or locations.\\
    \hline
    Material & Difference between goal and affordances - Exploring or responding to the properties and possibilities of tools and media.\\
    \hline
    Semiotic & Difference between sign and object - Shifting meanings and interpretations of symbolic or aesthetic representations.\\
    \hline
    Temporal & Differences between past and future - Using time through rhythm and pacing, reflection, or iterative development. \\
    \hline
    \end{tabular}
    \label{tab:coding_scheme}
\end{table}

\section{Examples of Artistic Exploration}\label{section:interviews}
To validate our model and ground it in real-world practices, we conducted interviews with visual artists and art teachers. These participants were selected because they are experts in navigating perspectival dimensions to explore and exploit a creative possibility space, essential in artistic practice and creative pedagogy~\cite{gluaveanu2023difference,beghetto2007toward}. Our goal is to gain insight into how pluri-perspectivism functions in practice and to assess the relevance and completeness of our five-dimensional model. In addition, the interviews inform the design of schema-guided~\cite{lee2021dialogue,zhang2023sgp}, few-shot learning strategies~\cite{brown2020language} for a VLM-enhanced robot. These examples can be used to construct contextual prompts that reflect pluri-perspectivism and model real-world creative reasoning and dialogue.
We interviewed 10 visual artists (6 women, 4 men) 65 years of age and older, and 8 art teachers (5 women, 3 men) who offer drawing and painting courses for older adults. Four artists were recruited through a foundation that elects Artists of the Year in the Netherlands, while the rest were found via the authors' networks. All maintained an active visual arts practice across different regions in the Netherlands. Art teachers were recruited through art participation organizations and the authors' networks. Most interviews were conducted in person at participants' workspaces, except for one artist and one teacher, who were interviewed remotely.

In semi-structured interviews, we used an interview guide with topics to be covered (see Supplementary Materials), to ensure consistency across participants while supporting a natural conversation and flexible responses~\cite{patton2002qualitative}. In line with the goal of looking for examples of pluri-perspectivism, we investigated contextual exploration in artistic practices and creative teaching. We discussed work environments, creative processes, and collaborations. The interviews with art teachers also focused on pedagogical techniques to help others explore and exploit a creative possibility space. The interviews were conducted in person at the artists’ and art teachers’ working environments and lasted approximately 40 minutes. All participants gave informed consent for audio recording. Data were pseudonymized and digitally stored at the University. Transcriptions were generated using a locally run version of OpenAI’s Whisper~\cite{radford2022whisper} to ensure privacy.

Thematic analysis using ATLAS.ti~\cite{atlasti2024} was guided by our five-dimensional model, as shown in Figure~\ref{fig:pentagram}. A coding scheme was developed with the perspectival dimensions as primary categories, defined in Table~\ref{tab:coding_scheme}. An elaborate coding scheme including subcategories can be found in Supplementary Materials. The use of subcodes in the qualitative coding process enabled a more nuanced and structured categorization of the data. While main codes captured broader thematic areas, subcodes allowed for the differentiation of specific patterns within those areas, enhancing the depth and clarity of analysis. This hierarchical approach aligns with best practices in thematic analysis~\cite{patton2014qualitative}. We validated outcomes through researcher triangulation, with independent coding by the first and last author. Intercoder reliability was assessed using Kippendorf’s c-a-binary ($\alpha$ = 0.549). The moderate agreement observed can be explained by the theoretical foundation of pluri\hyp{}perspectivism. As the model describes a multidimensional possibility space, it does not consist of discrete or exclusive categories. This can be expected to affect intercoder agreement. Overall, agreement was found in code selection, with variation mainly in segment granularity.

\subsection{Pluri\hyp{}perspectivism in Artistic Practices}
Qualitative analysis was guided by five pre-defined perspectival dimensions: Social, Semiotic, Temporal, Material, and Spatial. While no additional main categories were found, data revealed subcategories per dimension, and provided valuable insights into the manifestations of artistic contextual exploration. Both artists and art teachers engaged with all five dimensions, but differed in how they applied and emphasized them. For example, one oil painter described slow, deliberate work, often taking incubation time, while another oil painter pushed material boundaries by continuously layering oil paint until the artwork seemed to \textit{``take on a life of its own.''} Likewise, one art teacher emphasized free body movement to encourage expressivity, while another took a more analytical approach, having students repeatedly explore a particular topic, while constraining the use of materials. Below, we provide an overview of the most commonly described actions, grouped by the five perspectival dimensions.

\subsubsection{Social Perspectives}---Both artists (n=8) and art teachers (n=8) emphasized taking others’ viewpoints to gain new insights—primarily through conversation, observation, and collaboration. \underline{Talking about work} was central: artists~(n=6) discussed with peers, while teachers~(n=6) supported students in articulating their goals. \underline{Observing and studying art} was also common—mentioned by artists~(n=6) and teachers~(n=5)—alongside background research on thematic content. Artists (n=6) mentioned \underline{peer collaboration} through co-creation, exhibitions, and shared studios. All teachers~(n=8) organized \underline{group debriefings} to reflect on diverse approaches and encourage artistic freedom. As one teacher noted: \textit{``It makes them realize how many differences you see. And that gives a really nice feeling—the sense of all the different possibilities.''} Teachers~(n=6) also described group dynamics, such as mirroring and contagion. While all artists valued social exchange, 6 preferred working alone and not sharing initial ideas—suggesting that the need for social exchange may vary across creative stages.

For the art teachers, the social perspective involved their role in the classroom. All eight emphasized a coaching approach that involved stimulating creativity, offering guidance, and fostering personal growth. As one teacher put it, \textit{``Self-confidence and personal growth… that's what it's all about.''} All teachers mentioned using educational techniques such as scaffolding and demonstration, based on their own artistic expertise. In response to observed needs of students, teachers~(n=4) acted as disruptors to challenge students, or stepped back to observe~(n=4), leaving space for self-directed exploration.

\subsubsection{Semiotic Perspectives}---Artists (n=10) engaged in creative dialogues involving signs and meaning, while art teachers (n=7) supported such exchanges with their students. All artists~(n=10) described their process as a \underline{reflective dialogue}, an ongoing interplay between manifestation and imagination, guiding artistic direction. Several artists~(n=5) emphasized authenticity, and staying true to personal artistic goals. \underline{Conceptual exploration} was equally important across groups: artists~(n=9) and teachers~(n=6) described engaging in or encouraging experimentation through recombination, variation, and transformation. Teachers often used structured challenges to promote this. \underline{Surprise and serendipity} were also central to creativity—highlighted by 8 artists and 4 teachers—as vital for exploration. As one teacher noted: \textit{``The assignment is a kind of surprise egg. It always leads to something new.''} 

\begin{figure}[t] 
\centering
\includegraphics[width=\linewidth]{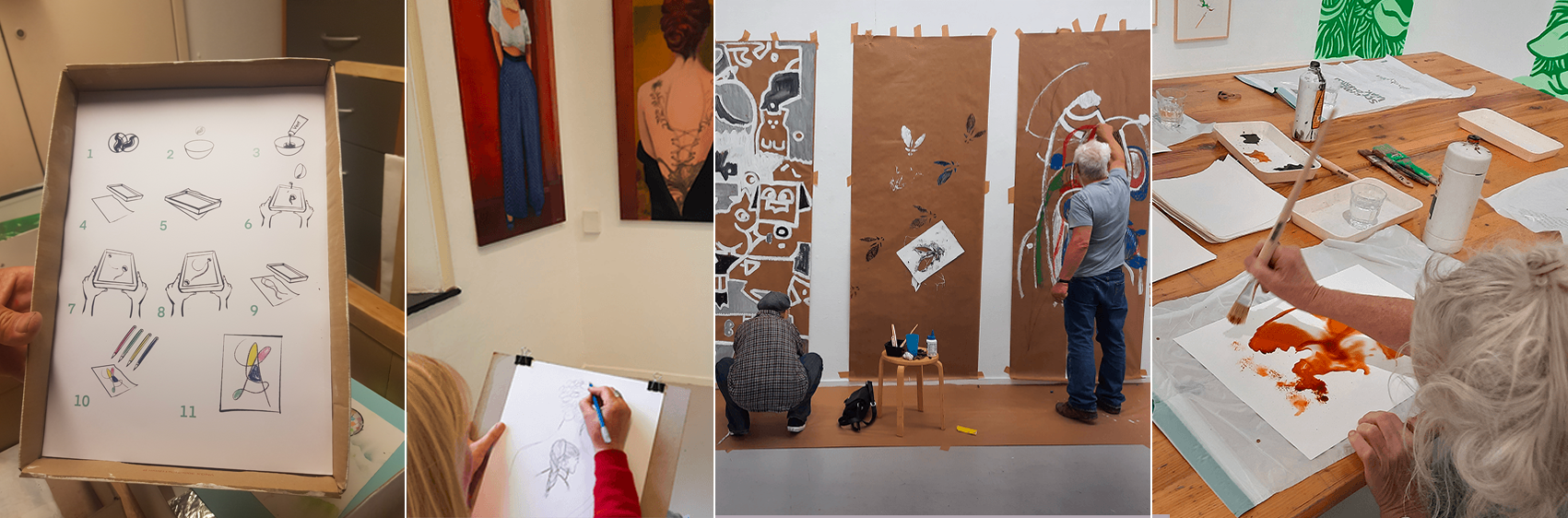} 
\caption{Impression of art teacher approaches i.e., guided instructions; working with art in the environment; promoting bodily motion; working with restrictions and unpredictable materials. Photos by the first author and by WG-Kunst, with permission.}
\label{fig:teacher-interviews}
\end{figure}

\subsubsection{Material Perspectives}---Artists~(n=10) aimed to push artistic boundaries by trying out \underline{different materials and techniques}, while art teachers~(n=8) actively promoted this with their students. Both groups emphasized the importance of engaging with materials and responding to their properties. These \underline{dialogues with} \underline{materials} allowed for intuitive, responsive interaction based on what the material seemed to `want.'' One artist, working with tubes, explained: \textit{``If you hang around with tubes long enough, the tubes start talking to you... What do those tubes actually want?''} In these material dialogues, artists~(n=6) and teachers~(n=3) looked for \underline{unexpectedness} to spark creative breakthroughs, and mentioned the use of unpredictable material behavior, inspiring new directions and further exploration. In addition, teachers sometimes introduced \underline{constraints}~(n=6), limiting tools or materials, to encourage students to push their creative boundaries by tackling well-defined challenges. Rather than restricting creativity, these constraints were seen as valuable prompts for exploration (Figure~\ref{fig:teacher-interviews}).

\subsubsection{Spatial Perspectives}---Changing spatial viewpoints emerged as an important strategy for gaining fresh perspectives. Both artists (n=10) and teachers (n=6) recognized the value of \underline{shifting viewpoints}, encouraging practices like stepping back or changing positions to see work from a different angle. Artists~(n=5) and teachers~(n=4) described how such shifts could reveal new insights and improve the creative process. As one artist put it, \textit{``You walk around and search for that spot, the place where the image speaks to you.''} In addition to adjusting viewpoints, artists~(n=7) reported on \underline{changing work locations}. Some artists~(n=4) described working in their studio and in nature: "\textit{``In the winter, I work here, but in the summer, well, then I wake up in the morning with a brilliant idea, and I go to the beach to try and build or realize it.''} Teachers~(n=3) also mentioned the importance of an inspirational work environment, for example organizing art classes in an exhibition space. Spatial perspectives were also explored through \underline{body movement}, allowing for multisensory creative experiences. Artists experimented with shifting from very small to very big canvas sizes~(n=4), while teachers sometimes encouraged bodily movement (Figure~\ref{fig:teacher-interviews}) to promote expressivity~(n=2).

\subsubsection{Temporal Perspectives}---Time was often mentioned by artists (n=10) and teachers (n=8). For artists, time was a flexible resource that allowed ideas to evolve. Artists~(n=6) described \underline{incubation} as essential for gaining insights: \textit{''I can sleep on it, and the next morning, I see it clearly.''} In contrast, art teachers focused on helping students navigate \underline{time constraints} within class settings, e.g., using structured assignments that support getting started and keep going~(n=8): \textit{''It's really an appetizer to get you started.''} Artists spoke of looking back at previous work~(n=6), sometimes reusing or refining earlier pieces, and used \underline{recurring themes}~(n=8) as a way to deepen exploration. Recurrence and reflection were often mentioned as contributing to \underline{artistic development} over time, highlighted by 7 artists. Teachers~(n=6) also aimed to promote artistic growth by offering adaptive challenges that evolved with students’ goals and competenties.

\section{Discussion}\label{section:discussion}
In this study, we introduced a model of pluri-perspectivism as a contextual map for human–robot co-creativity. Building on Glăveanu and Gillespie's concept of creatogenetic differences~\cite{gluaveanu2014creativity}, the model integrates five domain-specific dimensions into a layered structure distinguishing between collaborative (outer layer) and creative-task (inner layer) interactions. This framework captures how shifts across dimensions—such as a spatial re-orientation leading to a semiotic reinterpretation—can generate creative openings.

\subsubsection{Empirical Insights}---The findings of this study showed how pluri\hyp{}perspectivism is ubiquitous in creative practice. Participants described engaging with al five dimensions in personal and situated ways. They mentioned making perspectival shifts, such as using bodily movement (spatial) to inspire expressive marks (semiotic), or allowing time (temporal) for conceptual redirection. The examples supported the relevance of a pluri-perspectivist framework for understanding exploration in creative contexts.

\subsubsection{Implications for HRI}---The pluri-perspectivist model serves as a conceptual tool for understanding how a co-creative possibility space can be explored in different dimensions. This study contributes to a better grounded understanding of what context sensitivity entails in human-robot creative collaboration, and informs implementation strategies for HRI with VLM-enhanced robots. The model also provides insights into the challenges when using VLM-enhanced robots as co-creative agents. As described in Section~\ref{section:related-works}, social and semiotic dimensions may align well with schema-guided, few-shot prompting~\cite{lee2021dialogue,zhang2023sgp,brown2020language}, while temporal and spatial dimensions cannot be addressed through prompting alone. Additional modules can be implemented for this. Embodied co-creativity remains challenging, but VLM-enhanced robots can contribute uniquely through precision, rapid iteration, non-human perception and exploration, offering novel forms of creativity support. Furthermore, conversational perspective sharing can stimulate creativity in humans by prompting reflection, reinterpretation, and reframing. We will investigate this in our future co-creative HRI studies.

\subsubsection{Limitations and Future Directions}---While our pluri-perspectivist model offers a theoretically grounded framework for contextual understanding in creative collaboration, its application in HRI remains exploratory. As such, we view it as a conceptual contribution, that can inform human-robot interaction design, in-context learning, and the evaluation of perspectival diversity in co-creative interactions. Our future work will involve simplified implementations of the model with VLM-enhanced robots in constrained HRI scenarios, to assess how prompting different dimensions and perspectival shifts can support creative experiences in human-robot co-creativity with older adults.

\section{Conclusion}\label{section:conclusion}
We introduce a five-dimensional model of pluri\hyp{}perspectivism, grounded in creativity theory. Through insights from artists and art teachers, we demonstrate that pluri\hyp{}perspectivism is ubiquitous in creative practices. Literature study showed that the model could serve as a contextual map that structures interactions and be used for schema-guided prompting with a VLM-enhanced co-creative robot. The model can also serve as a tool for analyzing co-creative human–robot interactions. While the model has not yet been tested in human-robot interactions, we identify opportunities for its application in future studies. The proposition sets the stage for further research into fostering context-sensitive robots to support human creativity.

\begin{credits}
\subsubsection{\ackname} This work is part of the project Social Robotics and Generative AI to support Creative Experiences for Older Adults (NWO Doctoral Grant for Teachers, project no. 023.019.021).

\subsubsection{\discintname}
The authors have no competing interests to declare that are relevant to the content of this article. 
\end{credits}

\bibliographystyle{splncs04}
\bibliography{articles}

\end{document}